# INTRUSION DETECTION MECHANISM USING FUZZY RULE INTERPOLATION

[1]MOHAMMAD ALMSEIDIN, [2]SZILVESZTER KOVACS

[1, 2] Department of Information Technology, University of Miskolc, H-3515 Miskolc, Hungary

E-mail: [1] alsaudi@iit.uni-miskolc.hu, [2] szkovacs@iit.uni-miskolc.hu

## ABSTRACT

Fuzzy Rule Interpolation (FRI) methods can serve deducible (interpolated) conclusions even in case if some situations are not explicitly defined in a fuzzy rule based knowledge representation. This property can be beneficial in partial heuristically solved applications; there the efficiency of expert knowledge representation is mixed with the precision of machine learning methods. The goal of this paper is to introduce the benefits of FRI in the Intrusion Detection Systems (IDS) application area, in the design and implementation of the detection mechanism for Distributed Denial of Service (DDOS) attacks. In the example of the paper as a test-bed environment an open source DDOS dataset and the General Public License (GNU) FRI Toolbox was applied. The performance of the FRI-IDS example application is compared to other common classification algorithms used for detecting DDOS attacks on the same open source test-bed environment. According to the results, the overall detection rate of the FRI-IDS is in pair with other methods. On the example dataset it outperforms the detection rate of the support vector machine algorithm, whereas other algorithms (neural network, random forest and decision tree) recorded lightly higher detection rate. Consequently, the FRI inference system could be a suitable approach to be implemented as a detection mechanism for IDS; it effectively decreases the false positive rate value. Moreover, because of its fuzzy rule base knowledge representation nature, it can easily adapt expert knowledge, and also be-suitable for predicting the level of degree for threat possibility.

## 1. INTRODUCTION

Regarding to the rapid development of technology, the number of intrusions increased and developed continuously. In every day, there are large amounts of financial loss, privacy violation and information transfers in an illegal way as a result of succeeding intrusions implementation. There are different types of intrusions threatening network, computer information and resources. Many types of intrusions exist, such as user to root intrusion, where the goal of this type of intrusion is to have a full right permission of computer and network resources. Probing intrusion is another type of intrusion where the goal is to determine the weaknesses of computer and network resources based on scanning techniques. The previous types of intrusions could be implemented to be prerequisite steps of the Denial Of Service (DOS) intrusions. This type of intrusion heading to consume various resources in order to close-down several services for legal users. The distribution of DOS attacks represents the 60% of total number of attacks around the world [1].

IDS is one of the effective solutions to detect and prevent intrusions occurrence. According to the large amount of financial loss and privacy violation of intrusions, IDS has become a fundamental solution of network security. There are different challenges in implementing a sufficient IDS. One of these challenges is the binary decision in detection techniques. The typical detection mechanisms of IDS had a boundary problem [2]. The fuzzy system offers several advantages to handle the boundary problems. It also presents the detection degree level of intrusions which could be more readable for the security engineer. In this work, we break down the implementation of FRI inference system as a detection mechanism for DDOS attacks into three main steps.

- Step 1: To identify observable features suitable for IDS and the way they can handle the intrusion boundaries problem.

- Step 2: To implement FRI-IDS model as a detection mechanism for DDOS attacks.
- Step 3: To compare the FRI-IDS model with other literature's results, which had used the same test-bed environment with different classification algorithms for detecting DDOS attacks.

The rest of paper organized as follows: section (2) presents recent works related to the application of the fuzzy system for IDS. In section (3) typical types of IDS are introduced. Distributed denial of service attacks is illustrated in section (4) then the fuzzy rule interpolation is introduced in section (5). The analysis of DDOS dataset and preprocess steps illustrated in section (6). In section (7) the implementation of FRI-IDS model as a detection mechanism is introduced in details, followed by experiments and discussion in sections (8). The difference of prior works is introduced in section (9). Finally, section (10) concludes the paper.

## 2. IDS FOR DDOS ATTACKS

This section presents some relevant works related to the application of fuzzy system for intrusion detection. It also provides a brief overview of different methods and approaches that are used for intrusion detection.

In [3], authors implement an architecture to detect DDOS attack using Fuzzy Reasoning Spiking Neural-P (FRSN-P). It is a type of membrane computing system. The neurons within this system communicates based on electrical spikes (impulses). The Knowledge Discovery Databases (KDD-99) dataset imported in the proposed system. KDD-99 dataset was prepared by the University of California [4]. The constraint was on a synchronization flood. Authors extracted the synchronization flood attack from KDD-99 dataset. After extracting the required records of the desired attack, the fuzzy reasoning spiking neural-P was implemented and evaluated. According to the presented results, the proposed system was able to reach 0.02% false negative and 0.25% false positive detection rate.

Authors in [5], proposed a network IDS based on automatic fuzzy rule base generation. The single length of frequent item approach was implemented as preprocess step to generate the required fuzzy rule base automatically. KDD-99 dataset was used to evaluate the proposed system. The implemented experiments demonstrated that the proposed system obtained 90% as overall accuracy rate.

The work of [6], focuses on detecting and preventing the Neptune attack. It is a type of TCP flood attack and belongs to DOS attacks. The enhanced release of KDD99 dataset (NSL-KDD) was imported to test and evaluate the proposed fuzzy system [7]. According to the implementation of the proposed fuzzy system, feature ranking algorithm was implemented to select the relevant features for the detection approach. The proposed fuzzy system was evaluated with decision tree algorithm using the NSL-KDD dataset. It was succeeded to obtain 0.93% as the overall average accuracy rate for detecting Neptune attack. Compared with decision tree algorithm, the proposed fuzzy system had the highest average accuracy rate.

There are many IDS methods implementing feature selection algorithms. A large number of features can be collected using several network tools. However, not all of them are relevant to the detection mechanism. The primary aim of feature selection algorithms is to reduce the computation time by reducing the size of data. It can reduce the problem space by reducing the required feature set to a minimum [8].

Authors in [2], proposed a network IDS based on fuzzy system. Feature selection algorithm was applied for reducing the originally large (41) number of features of the KDD-99 dataset. The authors divided the imported dataset into two parts, the first part for training phase and the second part for testing phase. The fuzzy rules generated based on the instructions of a knowledge expert. The proposed fuzzy system was able to achieve 0.95% average accuracy. Other algorithms were also used in combination with a fuzzy system to detect and prevent intrusions. In [9], a hybrid approach of genetic algorithm and fuzzy system was proposed for detecting DOS attacks. The main purpose of implementing genetic algorithm was the automatic fuzzy rule generation, as a preprocess step of the IDS construction. The testbed environment was the KDD-99 dataset. The proposed system achieved 0.94% average detection rate.

There are other hybrid fuzzy IDS solutions too. In [10], the authors combine the neural network and fuzzy system to enhance the detecting rate of intrusion detection. The fuzzy rule base is generated based on expert knowledge base. The neuro fuzzy IDS achieved 0.93% average detection rate. In [11], a combination of fuzzy system and decision tree algorithm was implemented and evaluated. Features selection was applied to reduce the size of feature space. Moreover, the decision tree algorithm was applied to extract automatically the fuzzy rule base. The proposed system reached 0.99% average accuracy rate.

In [12], authors focused on enhancing the detection rate of IDS based on combining the fuzzy system and Particle Swarm Optimization (PSO) method. The PSO was applied for generating fuzzy rule base in order to detect DDOS attacks. As a result of the PSO generated fuzzy rules, the proposed fuzzy system was able to reach 0.93% as a detection rate average.

The past works provided convincing contributions and supporting the idea that the fuzzy rule based model is a useful device for IDS implementation. The goal of this paper is to introduce the benefits of the FRI application at the IDS application area, mainly tackling the distributed denial of service type attacks. The aim behind using the FRI is the simplification of the expert rule base and the extension of the binary decision problem to continuous truth value, in which conclusions like "the level of the attack" can be also simply defined.

## 3. IDS IMPLEMENTATION

This section provides a brief overview to the IDS problem domain. It also shortly describes how IDS detects intrusion within the network. IDS is one of the significant solution that's monitors the network traffic, in order to observe and detect intrusions. Recently, IDS has become a fundamental component in network security design. Any attempt to use and consume network/computer resources is presented as DOS intrusion. One of the main functions of IDS system is to generate alerts when an intruder appears within the protected network. IDS can be implemented either as a software platform or as a hardware device.

Typically, IDS includes the following three modules: audited module, analyzer module and response module [13]. There are different types of IDS that can be distinguished depending on the source of audited data within network and also based on the implemented detection mechanism.

According to the applied detection mechanism, there are two types of IDS [14, 15]:
- Anomaly based detection
- Signature based detection

The signature-based detection implements the search mechanism for known sequence series of packets (signature) of intrusion. If the current series of packets/bytes matched the stored sequence series of packets, the intrusion alert raised. It is extremely efficient with no false alarms for past detected intrusions. On the other hand, it had weaknesses, such as it is only efficient with the known intrusions that had a stored signature [16].

The anomaly-based IDS detects the intrusions based on comparing the current network traffic with the historical baseline data of normal traffic. If the current network traffic exceeds the predefined baseline normal traffic, the intrusion alert raises. It had a chance to detect the novel intrusions with a high proportion of false positive alerts. Additionally, anomaly based detection mechanism requires a large amount of training data for identifying the normal behavior [17].

IDS can be also categorized based on the source of audited data as follows: Network Intrusion Detection System (NIDS) or Host Intrusion Detection System (HIDS). HIDS monitors and protects a specific device within the network. Meanwhile, NIDS implemented to protect and detect intrusions of all the connected devices within a network.

## 4. DDOS ATTACKS

The DDOS attack is treated as one of the most harmful types of attack. Nowadays, DDOS attacks are considered as a continuous challenge for both users and organizations. The first serious DDOS attack was appeared in 2000 against Yahoo [18]. The main purpose of DDOS attacks is to consume different types of resources such as network bandwidth, CPU, memory utilization etc. Any consumption of these resources will increase the overloading and as a result different services would be unavailable for legal users.

In 2015, according to Kaspersky security report [19], the approximate cost of DDOS attack were 52000 $ for small businesses and around 440,000 $ for enterprise businesses. In 2016, the dangerous effect of DDOS reached 80 countries around the world [20]. There different types of (DDOS) attacks such as smurf attack where an intruder sends large numbers of Internet Control Message Protocol (ICMP) echo packets to the intended victim. In most situations the intermediary (slave) machine does not filter ICMP messages, therefore, many clients on the network who receive this ICMP echo request send ICMP replay back. Another type of DDOS attacks is the User Datagram Protocol (UDP) flood attack. It is one of the most common types of DDOS attack, where the intruders send large number of UDP traffics to the victim within a specific period of time. From another perspective, the HTTP-flood attack is considered as a difficult one to detect. According to HTTP-flood attack, the intruder sends completely normal posted messages with a very slow rate in a systematic way, this type of DDOS is difficult to detect because its behavior seems as a normal behavior [21].

Another type of modern DDOS attack is a Simple Query Language (SQL) Injection Distributed Denial Of Service (SIDDOS). According to this type of DDOS, the intruder inserts a malicious SQL statement as a string in the browser side, then it is forwarded to the victim as an executed statement [22].

## 5. FUZZY INFERENCE SYSTEM

The term of fuzzy logic was produced by Professor Lotfi Zadeh [23]. Fuzzy logic appears in many successful sophisticated systems in many application areas. There are some application areas where the two valued logic and the related binary decision could lead to inefficient solutions. Fuzzy logic offers several advantages to handle the binary decision problems [24].

There are some requirements emerging during the design and implementation of fuzzy system. The inputs and outputs of the fuzzy system should be clearly defined, then the fuzzy partitions of the input and output universes should be established, then the fuzzy rule base must be completely prepared.

Fuzzy partitions of the input values provide a significant way to define the real input value with each predefined linguistic term [25]. For generating a conclusion by a fuzzy system, first the crisp inputs are transformed to fuzzy sets by the fuzzifier, then from the fuzzified input, the fuzzy inference system calculates the fuzzy conclusion. At the end, the crisp output is prepared by defuzzification of the fuzzy conclusion [26].

In typical classical fuzzy inference system, the fuzzy rule base is extracted from expert knowledge. To be able to handle all the possible input values, the fuzzy rule base must cover all the input universes. Therefore, the step of the fuzzy rule base considered as the most critical step during the design of the fuzzy system. In general, generating a complete fuzzy rule base in a multidimensional problem is difficult to be implemented because of the lack of information for all the possible fuzzy rules. In case of missing rule definitions, there could be some observation which is not covered by any of the fuzzy rules, in that case no conclusion can be gained from the rule base. The FRI methods can solve this situation [27]. FRI methods can generate their approximate conclusions either directly from inputs by an interpolating fuzzy function, or by interpolating a new fuzzy rule which overlaps the input [28].

## 6. DDOS DATA-SET

There are several open source datasets that exist which include intrusions related data. These datasets provide a convenient environment for research purposes. In this work, the DDOS dataset of [21] was used as a test-bed environment for testing the FRI inference based IDS solutions.

The DDOS dataset includes intrusions related data, such as HTTP flood and SIDDOS. This dataset can be downloaded freely for research purposes from [29]. The distribution of the recorded attack types within the DDOS dataset summarized in Table 1.

*Table 1: Distribution Of DDOS Dataset Attacks.*

| Attack | Number of Records |
|---|---|
| SIDDOS | 6665 |
| HTTP Flood | 4110 |
| UDP Flood | 201344 |
| SMURF | 12590 |

The discrete and continues features appearing in the DDOS dataset are presented in Table 2 and Table 3 respectively.

*Table 2: The Discrete Features Of DDOS Dataset.*

| Index | Features | Description |
|---|---|---|
| 6 | PKT TYPE | Packet Type Based on Used Protocol |
| 8 | FLAGS | 7-Digit Flag Strings |

| 11 | NODE NAME FROM | Client Name That Sends The Packet |
|---|---|---|
| 12 | NODE NAME TO | Client Name That Receives The Packet |
| 28 | PKT CLASS | The Class of Packet |

*Table 3: The Continues Features of DDOS Dataset.*

| Index | Features | Description |
|---|---|---|
| 1 | SRC ADD | Port of Source Address |
| 2 | DES ADD | Port Destination Address |
| 3 | PKT ID | Packet Identifier |
| 4 | FROM NODE | Define client sending packet |
| 5 | TO NODE | Define client receiving packet |
| 7 | PKT SIZE | The Packet Size in bytes |
| 9 | FID | Flow Identifier |
| 10 | SEQ NUMBER | Sequence Number |
| 13 | NUMBER OF PKT | Total Number of Packets |
| 14 | NUMBER OF BYTE | Total Number of bytes |
| 15 | PKT IN | Total Time of Packet Inside Queue |
| 16 | PKT OUT | Total Time of Packet Outside Queue |
| 17 | PKT R | Time of Packet Received |
| 18 | PKT DELAY NODE | TimePacketDelayWithin Node |
| 19 | PKT RATE | Average Packet Rate |
| 20 | BYTE RATE | Average byte Rate |
| 21 | PKT AVG SIZE | Average Packet Size |
| 22 | UTILIZATION | Bandwidth Utilization |
| 23 | PKT DELAY | Total Time of Packet Delay |
| 24 | PKT SEND TIME | Time of Sending Packet |
| 25 | PKT RECEIVED TIME | Time of Receiving Packet |
| 26 | FIRST PKT SENT | Time of First Packet Sent |
| 27 | LASTPKT RECEIVED | Time of Last Packet Received |

### 6.1 Dataset Preprocessing And Features Selection

The DDOS dataset consisted of a large number of connection records. Because of that, we extracted 10% of total number of intrusions records. The extracted DDOS dataset listed in Table 4.

*Table 4: The Extracted DDOS Dataset.*

| Class Name | Number of Connection Records |
|---|---|
| SIDDOS | 676 |
| HTTP Flood | 441 |
| UDP Flood | 20135 |
| SMURF | 1260 |
| Normal | 193000 |

The typical IDS detects the packet based on predefined rules such as SNORT [30]. The rules which are responsible for distinguishing intrusion from normal packets must be created based on the features given in the intrusion dataset. There is a large number of features could be recorded during the collection of intrusions dataset. These features could be recorded using any network monitoring tools. Generally, most of them are not relevant features. It means that features are not relevant in the detection of a given type of intrusion. Features selection considered as an important step because if there are irrelevant features then it could decrease the performance of the final IDS.

Features selection in the example of this paper the Information Gain (IG) algorithm [31] was selected. The IG algorithm is based on the concept of entropy. According to [32], entropy parameter computed to characterize the relevantly of each feature. Suppose that, E = (E,P) be a discrete probability space, where E = $\{E_1, E_2, ...., E_n\}$ is the finite set of the selected features. Each of selected features had the following probabilities $P_i$, i =1, 2,....., n. The entropy for each features computed as Equation 1 illustrated.

$$\text{Entropy}(E) = -\sum_i p_i \log_2 (p_i) \quad (1)$$

After the entropy parameters calculated for each feature of the DDOS dataset, the next step is to calculate the IG values. IG computed based on the predefined collected entropy parameters and the set of all possible values for the feature, Equation 2 presents the calculation formula of IG.

$$IG = \text{Entropy}(E) - \sum \frac{E_K}{n} \text{Entropy}(E_k) \quad (2)$$

Where n presents the total number of instance of records, $E_k$ denotes the total number of instance of records that belongs to the class k. Table 5 summarizes the top ten relevant features using IG algorithm.

*Table 5: The Relevant Features Using IG Algorithm.*

| No. | Features | IG Values |
|---|---|---|
| 1 | PKT RATE | 0.3811300 |
| 2 | BYTE RATE | 0.3809683 |
| 3 | UTILIZATION | 0.3809683 |
| 4 | PKT SIZE | 0.3803146 |
| 5 | PKT AVG SIZE | 0.3786557 |
| 6 | NUMBER OF PKT | 0.3740723 |

| 7 | PKT DELAY | 0.3630512 |
|---|---|---|
| 8 | NUMBER OF BYTE | 0.3630512 |
| 9 | FIRST PKT SENT | 0.3513514 |
| 10 | PKT DELAY NODE | 0.3321921 |

## 7. FRI-IDS MODEL GENERATION

In this section the full architecture of FRI-IDS model discussed including its main functions and interactions. Typically, in classical fuzzy system, the inferring of consequences could not be deduced in case if some situations were not explicitly defined in a fuzzy rule based. Therefore, the inferring of consequences of fuzzy system required a completed fuzzy rule base.

In case of sparse fuzzy rules which were not covered all of possible situations, FRI methods offer the capability to generate the possible inference, even in case of lack definitions and information of existing knowledge representation. This benefit could be beneficial in partial heuristically solved applications. Suppose that, there is an observation x existed and did not explicitly defined in a fuzzy rule based.

Equation 3 presents the union of the antecedent part of fuzzy rules in sparse rule [28].

$$\bigcup_{i=1}^{k} supp(A_i) \subset X \rightarrow \bigcup_{i=1}^{k} supp(A_i) = 0 \quad (3)$$

In the opposite case, if there is a completed fuzzy rule base, then the union of an antecedent part of fuzzy rules was covering all of the input universes as Equation 4 presented.

$$\bigcup_{i=1}^{k} supp(A_i) = X \quad (4)$$

*supp* refers to support. The support of a fuzzy set indicates the set of all elements within the universe of discourse which their degree of membership is greater than zero. Further, $X_i$ is the $i_{th}$ input universe of discourse, $A_{ik}$ is the $k_{th}$ set of the partition of $X_i$. Additionally, Figure 1 illustrates the case of sparse rule, when the observation x appeared and was not covered by any fuzzy rule base. Figure 2 illustrates the case of complete rules, when the observation x appeared and was covered by fuzzy rule base.

The FRI-IDS model was a description of the problem domain (IDS application area). It constraints of the key features (relevant features) to detect the intrusion.

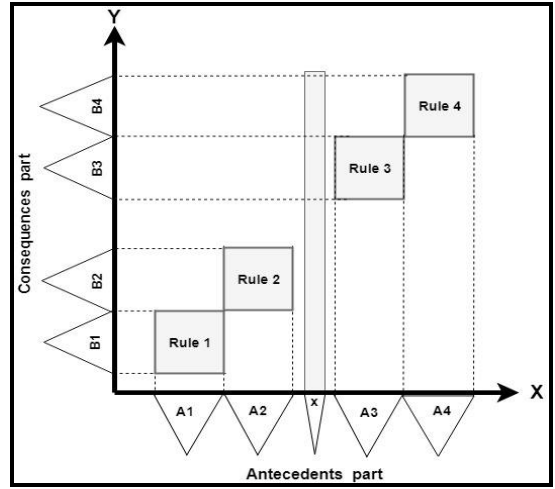

*Figure 1: The Case Of Sparse Fuzzy Rule*

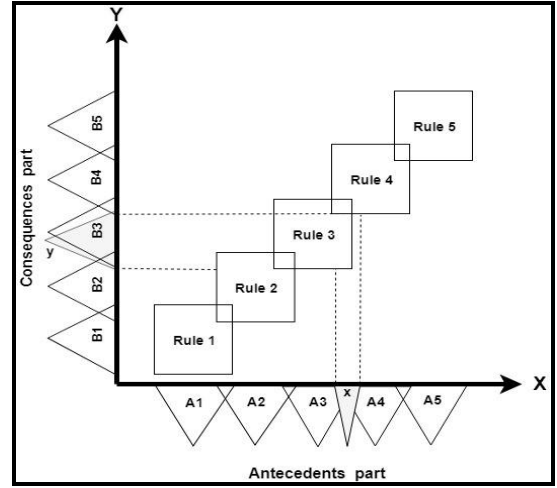

*Figure 2: The Case Of Complete Fuzzy Rules*

There are four major components needed to implement the FRI-IDS model as a detection mechanism:

- Setup the input and output of the FRI-IDS model. (The input parameters of FRI-IDS model were the relevant features of the test-bed dataset which are mentioned in Table 5, the output of FRI-IDS model supposed to be the level of attack instead of binary decision).

- Setup the fuzzy sets for each input/output of the FRI-IDS model. (This component introduced in details in section (7.2)).

- Setup the fuzzy rules for all the possible events of normal and intrusion. (This component introduced in details in section (7.2)).

- Testing and validating the FRI-IDS model.

The inference engine of the FRI-IDS model was performed by the Fuzzy Interpolation based on the Vague Environment method (FIVE). It was introduced by [33, 34, 35] in 1996. The FIVE method serves the deducible conclusions even in case if some situations are not explicitly defined in a fuzzy rule based knowledge representation. It is produced to serve many application areas such as IDS solution, which is served a crisp observation and at the same time required a crisp conclusion. It is worth mentioning that since using the FRI (FIVE) method as a inference engine there is no need for an additional defuzzification step.

The architecture of the FRI-IDS model was shown in Figure 3 starts by data filtration phase where the network traffics (training data) analyzed in order to extract and determine the relevant features. During the data filtration phase, the irrelevant features were removed. It should be known that, the existence of irrelevant features could decrease the performance of FRI-IDS model. In the modeling phase the sparse fuzzy model identification [36] was performed, it had several actions, this includes the estimation of fuzzification and membership functions, fuzzy rules generation besides deducing the consequences and tuning methods.

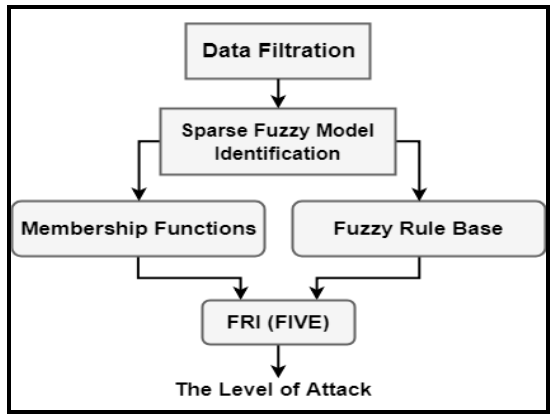

*Figure 3: The Architecture of The FRI-IDS Model*

### 7.1 Data Filtration Phase

The test-bed dataset was divided into training part and testing part. The training data consisted of 10000 records with 5000 normal cases and 5000 intrusion cases. The test data consisted of 10000 records with 5000 of normal cases and 5000 of intrusion cases. In order to increase the efficiency of IDS, it is important to identify the observable features that are relevant to detect intrusions from the network traffic data [3].

These are some of relevant features: utilization, packet rate, byte rate, pkt size and pkt delay. For the sake of reducing the possible number of fuzzy rules and low complexity system, the highest three relevant features according to the IG algorithm were used as input parameters of FRI-IDS model. These relevant features as found in Table 5 are the packet rate, byte rate and utilization. The anomaly based and misuse based detection techniques detect the attacks based on the predefined rule base (i.e. rules for normal and intrusions). For this, sorting the normal and intrusions cases of the training data is required [25]. Algorithm 1 presents the sorting and feature extraction of the training data.

| **Algorithm 1:** Sorting and Feature Extraction Algorithm |
|---|
| **Input**: The training data |
| **Output**: Two pools of test-bed dataset (normal and intrusion) |
| 1: **while** Termination Condition Not Met **do** |
| 2:    Classify whole test-bed dataset into "normal" and "attack" class |
| 3:    Check for missing entry for all records |
| 4:    Extract the suitable features for IDS based o IG algorithm |
| 5:    Remove all irrelevant features |
| 6:    Store the values of normal pool |
| 7:    Store the values of intrusions pool |
| 8: **end while** |

The outputs of sorting and feature extraction algorithm were two pools of normal and intrusion records. These two pools consisted with only the relevant observable features that are suitable for FRI-IDS model (packet rate, byte rate and utilization) features, where all other values are removed.

### 7.2 The Modeling Phase

The part of fuzzy modeling considered as one of the important parts in the fuzzy system. The FRI-IDS model was constructed by using the sparse fuzzy model identification [36]. The training data of FRI-IDS model had three input parameters (packet rate, byte rate and utilization). These parameters were chosen according to the IG algorithm to infer the DDOS attack. In order to generate the optimized fuzzy rules and fuzzy sets, the Rule Base Extension using Default Set Shapes (RBE-DSS) method [37] was applied. According to [37, 38] the main steps of RBE-DSS method can be summarized as follows:

- In the early stage of modeling the fuzzy system, the RBE-DSS method generates two rules that covered (fit) the minimum and maximum of the output.

- In the next step, the hill climbing tuning algorithm started. It is adjusting the previous parameter values one by one. For each iteration, the fuzzy system is evaluated with different parameters values based on training data. The retrieved parameters values ensure that the fuzzy system belongs to the better performance index for the later iterations.
- The performance index is computed in each iteration to compare the obtained results with different parameter values. The relative root main square error was chosen as a performance index for tuning the FRI-IDS model.
- On the assumption, the increasing of fuzzy system performance appeared too slow or interrupted (i.e. fuzzy system obtained the local minimum) then, the new fuzzy rule generated to increase the possibilities of fuzzy system enhancement.
- The new fuzzy rule created in where the difference between the value of actual output and computed output is the maximum.
- The tuning process stopped when the predefined performance index value obtained or when the number of iterations is reached.

As a result of applying the RBE-DSS method, Figure 4 shows the support of antecedent fuzzy sets of the tuned FRI-IDS model based on the training data.

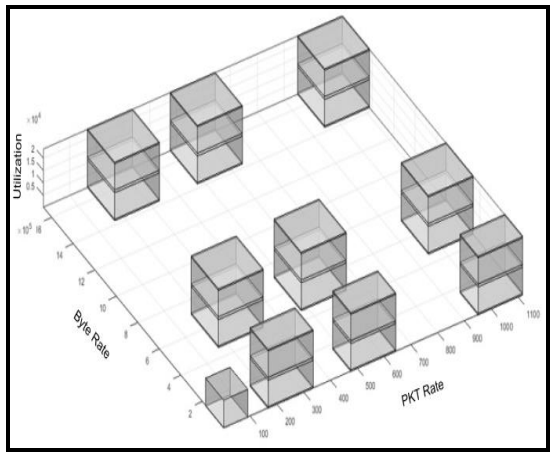

*Figure 4: Support of The Antecedent Fuzzy Sets of FRI-IDS Model*

It is worth mentioning that, the generated fuzzy rules by RBE-DSS method were sparse, these fuzzy rules for the fuzzy interpolation and if it is implemented for a classical fuzzy reasoning there is no result could be obtained. Out of 28 fuzzy rules were generated in order to detect the DDOS attack based on the training data. Table 6 presents the generated fuzzy rules•. Subsequently of modeling phase, the obtained fuzzy sets were represented by a trapezoidal membership functions. The byte rate and utilization input parameters have three trapezoidal membership functions and the packet rate input parameter has four trapezoidal membership functions. Table 7 (See Annexure 1) presents the optimized values of fuzzy sets for FRI-IDS model based on the training data.

*Table 6: The Obtained Fuzzy Rules.*

| No. | Packet Rate | Byte Rate | Utilization | Consequences |
|---|---|---|---|---|
| 1 | L | L | L | FA |
| 2 | L | L | M | FA |
| 3 | L | L | H | FA |
| 4 | L | M | L | FA |
| 5 | L | M | M | FA |
| 6 | L | M | H | FA |
| 7 | L | H | L | A |
| 8 | L | H | M | A |
| 9 | L | H | H | A |
| 10 | M | L | L | FA |
| 11 | M | L | M | FA |
| 12 | M | L | H | FA |
| 13 | M | M | L | FA |
| 14 | M | M | M | FA |
| 15 | M | M | H | FA |
| 16 | M | H | L | A |
| 17 | M | H | M | A |
| 18 | M | H | H | A |
| 19 | H | L | L | A |
| 20 | H | L | M | A |
| 21 | H | L | H | FA |
| 22 | H | M | L | A |
| 23 | H | M | M | A |
| 24 | H | M | H | A |
| 25 | H | H | L | A |
| 26 | H | H | M | A |
| 27 | H | H | H | A |
| 28 | VL | L | L | A |

FRI-IDS model serves crisp values and at the same time generates a crisp conclusion. Therefore, each observation within the training data in the example of this paper presented as a fuzzy singleton. Fuzzy systems had the capability to extension the binary decision to the continuous truth value which is more readable and easier to be understood and analyzed. Suppose that, there are two observations within the training data, the first observation had the following crisp values (packet

---

• HINT: VL = VERY LOW, L = LOW, M = MEDIUM, H = HIGH, FA = FALSE ATTACK, A = ATTACK.

rate= 200, byte rate = 55943 and utilization = 11560). The second observation had the following crisp values (packet rate= 900, byte rate = 1190251 and utilization = 22029). The inferred consequence of FRI-IDS model for the previous two observations was shown in Figure 5 and Figure 6 (See Annexure 2 and Annexure 3) respectively where the first observation presents the normal event and the second observation shows the DDOS attack event. FRI-IDS model can serve the interpolated conclusions even in case if some observations are not covered directly by fuzzy rules as Figure 6 (See Annexure 3) presented.

## 8. EXPERIMENTS AND DISCUSSION

This section illustrates the testing and validating of the FRI-IDS model using the test-bed dataset. Thereupon, all experiments were conducted using Matlab [26] and FRI toolbox [39]. The inference engine of FRI-IDS model was performed using FIVE method. The code of FIVE method and other FRI methods can be used through FRI toolbox which can be downloaded freely from [39]. The overall process of testing and validating the FRI-IDS model was shown in Figure 7.

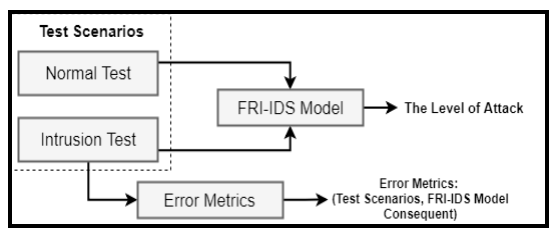

*Figure 7: The Testing and Validating Process of FRI-IDS Model*

The FRI-IDS model was tested and evaluated based on two test scenarios:

- The first test titled as the normal test scenario where 5000 instance of normal cases are used as input parameters of FRI-IDS model. The extracted 5000 instance of normal cases presented as matrix of normal test.

*Normal Test Matrix*

$$\begin{Bmatrix} 328 & 505434 & 23632 \\ 125 & 124944 & 5961 \\ 658 & 379060 & 3092 \\ . & . & . \\ . & . & . \\ \text{Pkt Rate} & \text{Byte Rate} & \text{Utilization} \end{Bmatrix}$$

- The second test scenario was titled as intrusions test scenario where 5000 instance of intrusion cases are used as input parameters of FRI-IDS model. The extracted 5000 instance of intrusion cases presented as a matrix of intrusion test.

*Intrusion Test Matrix*

$$\begin{Bmatrix} 1118 & 1677420 & 14377 \\ 963 & 1444030 & 12381 \\ 328 & 18067 & 845 \\ . & . & . \\ . & . & . \\ \text{Pkt Rate} & \text{Byte Rate} & \text{Utilization} \end{Bmatrix}$$

The previous two matrices of normal and intrusion test scenarios were chosen as a two input testing files of FRI-IDS model. The evaluation of FRI-IDS model carried through the computed error metrics of the test scenarios. The inferred consequence of FRI-IDS model was compared along with the actual values of normal and intrusion. According to [15], the error metrics of the test scenarios were extended to the following performance metrics:

- True Positive (TP): The total number of intrusion cases that inferred correctly by the FRI-IDS model.

- True Negative (TN): The total number of normal cases that inferred correctly by the FRI-IDS model.

- False Negative (FN): which is actually presented as important metric, the total number of intrusion cases that inferred incorrectly by the FRI-IDS model.

- False Positive (FP): the total number of normal cases that inferred incorrectly by the FRI-IDS model.

- Detection Rate (DR): is a performance metric of identifying the overall detection rate of the FRI-IDS model based on the test scenarios. It was computed by the total number of true positive cases and true negative cases divided by the total number of cases within the test scenario.

Furthermore, the previous mentioned performance metrics offer the capability to compare the FRI-IDS model results with other algorithms that have been implemented for detecting DDOS attacks. As a result of the test scenarios of FRI-IDS model, 10000 of cases were tested successfully. These cases were split as 5000 of normal cases and 5000 of intrusion cases. During the normal test scenario, only 3 records of normal cases were inferred incorrectly by FRI-IDS model. Besides, during the intrusion test scenario, 332 of intrusion cases were inferred incorrectly by the FRI-IDS model. The obtained results beside the error metrics values presented in Table 8.

*Table 8: The Result Of The Test Scenarios Cases.*

|  | Normal | Intrusion | Total |
|---|---|---|---|
| Normal | 4997 | 3 | 5000 |
| Intrusion | 332 | 4668 | 5000 |
| Total | 5329 | 4671 | 10000 |

According to the obtained results and the error metrics values of Table 8, the confusion matrix parameters of FRI-IDS model presented in Table 9.

*Table 9: Confusion Matrix Of FRI-IDS Model.*

| Alert Response | Intrusion Packet Prediction | Normal Packet Prediction |
|---|---|---|
| Intrusion | TPR = 0.93 | FNR = 0.06 |
| Normal | FPR = 0.0006 | TNR = 0.999 |

The implemented experiments have demonstrated that, the FRI-IDS model obtained 96.65% as an overall detection rate. The computed performance metrics concluded that, the FRI-IDS model obtained an acceptable value for the detection rate, and it decreases effectively the false positive rate. Decreasing the false positive rate helps to reduce the large amount of IDS alerts. To summarize the aforementioned results, the FRI-IDS model could be a suitable approach to be implemented as a detection mechanism for the following reasons:

- The FRI-IDS model offers extension of the binary decision problem to continuous truth value, in which the inferred consequence like "the level of intrusion", which makes the response result more readable and clearly analyzed rather than binary decision.

- It is difficult to identify a clear boundary between normal and intrusion packets. Therefore, fuzzy system effectively smooths the abrupt break of normal and intrusion.

- FRI methods can serve deducible (interpolated) conclusions even in case if some situations are not explicitly defined in a fuzzy rule based knowledge representation.

- The implemented experiments show that the FRI-IDS model obtained an accepted value for detection rate and false positive rate.

From another perspective, there are several convincing efforts of literature's to implement different classification algorithms in order to prevent DDOS attacks. Therefore, to support the idea of implementing FRI-IDS model as a detection mechanism could be a suitable approach. The obtained results of FRI-IDS model compared with Alkasassbeh [21] and Irfan Sofi [22] results. They have employed different classification algorithms to detect the DDOS attacks using the same test-bed environment. Table 10 summarized the comparison result of FRI-IDS model with other classification algorithms.

*Table 10: FRI-IDS Model Vs Classification Algorithms.*

| Authors | Algorithms | | | | FRI |
|---|---|---|---|---|---|
| Irfan Sofi et al. [22] | Neural Network | Naive Bayes | Decision Tree | Support Vector Machine | **FRI-IDS** |
| Detection Rate | 98.91% | 96.89% | 98.89% | 92.31% | **96.65%** |
| Alkasassbeh et al. [21] | Neural Network | Naive Bayes | Random Forest | | **FRI-IDS** |
| Detection Rate | 98.63% | 96.91% | 98.02% | | **96.65%** |

## 9. DIFFERENCE FROM PRIOR WORKS

According to the results, the overall detection rate of the FRI-IDS is in pair with other methods. On the example dataset it outperforms the detection rate of the support vector machine. FRI-IDS reduced effectively false positive rate value which reduced the large number of IDS false alerts. Moreover, FRI-IDS offers the extension of the binary decision problem to continuous truth value, in which conclusions like "the level of the attack" can be also simply defined. From another perspective, there are valuable efforts of implementing the classical fuzzy reasoning for intrusion detection as it mentioned previously in the section (2).

Nevertheless, these methods desire a dense fuzzy rules as a major requirement. Regarding to the large number of network connections, it could be very hard to comply the dense fuzzy rules. However, FRI-IDS model is characterized to offer the attack alert generation in case of lack of information and definition of existing knowledge base. It can generate their approximate conclusions either directly from inputs by an interpolating fuzzy function, or by interpolating a new fuzzy rule which overlaps the input.

## 10. CONCLUSION

This paper has investigated the capabilities to use the FRI methods in the IDS application area. This investigation is practiced by implementing the FRI-IDS model as a detection mechanism for DDOS attack. The FRI-IDS model was constructed using the sparse fuzzy model identification. The fuzzy rules of FRI-IDS model were generated and optimized using RBE-DSS method. In the example of the paper as a test-bed environment, an open source DDOS dataset was used. The implemented experiments have demonstrated that the FRI-IDS model obtained an accepted detection rate. It has reduced effectively the false positive rate value which decreased the large amount of IDS alerts. Additionally, the FRI-IDS model can serve the interpolated conclusions even in case if some observations are not covered directly by fuzzy rules.

The obtained results of FRI-IDS model compared with other literature's results which they employed different algorithms to detect the DDOS attacks using the same testbed environment, FRI-IDS model outperforms the detection rate of support vector machine algorithm in the example of DDOS dataset, where other algorithms (neural network, random forest and decision tree) recorded lightly higher detection rate. Consequently, the FRI-IDS model could be a suitable approach for detecting intrusions if it is implemented as a detection mechanism. It is characterized by offering the capability to present the detection level of intrusion and permits the attack alert generation in case of a lack of information and definition of existing knowledge base.

However, for future work, it could be useful to apply the FRI-IDS model along with other types of intrusions. Moreover, it is worthy, to examine how the FRI-IDS model could be applied with other FRI methods instead of FIVE method.

## ACKNOWLEDGEMENT

The described study was carried out as part of the EFOP-3.6.1-16-00011 "Younger and Renewing University – Innovative Knowledge City – institutional development of the University of Miskolc aiming at intelligent specialisation" project implemented in the framework of the Szechenyi 2020 program. The realization of this project is supported by the European Union, co-financed by the European Social Fund.## REFERENCES

[1] Y. Zhang, D. Zhao, and J. Liu, "The Application of Baum-Welch Algorithm In Multistep Attack", *The Scientific World Journal*, vol. 2014, Article ID 374260, 7 pages, 2014, DOI:10.1155/2014/374260

[2] R. Shanmugavadivu and N. Nagarajan, "Network intrusion detection system using fuzzy logic", *Indian Journal of Computer Science and Engineering (IJCSE)*, vol. 2, no. 1, pp. 101–111, 2011.

[3] R. K. Idowu, Z. A. Othman, et al., "Denial Of Service Attack Detection Using Trapezoidal Fuzzy Reasoning Spiking Neural P System", *Journal of Theoretical & Applied Information Technology*, vol. 75, no. 3, pp. 397-404, 2015.

[4] S. D. Bay, D. Kibler, M. J. Pazzani, and P. Smyth, "The UCI KDD Archive of Large Data Sets For Data Mining Research and Experimentation", *ACM SIGKDD Explorations Newsletter*, vol. 2, no. 2, pp. 81–85, 2000.

[5] S. P. Thakare and M. Ali, "Introducing Fuzzy Logic in Network Intrusion Detection System", *International Journal of Advanced Research in Computer Science*, vol. 3, no. 3, pp. 810-815, 2012.

[6] N. N. P. Mkuzangwe and F. V. Nelwamondo, "A fuzzy Logic Based Network Intrusion Detection System For Predicting The TCP SYN Flooding Attack", *In Asian Conference on Intelligent Information and Database Systems*, Springer, 2017, pp. 14–22.

[7] M. Tavallaee, E. Bagheri, W. Lu, and A. A. Ghorbani, "A detailed Analysis of the KDD CUP 99 Dataset", *In Computational Intelligence for Security and Defense Applications. CISDA. IEEE Symposium on*, IEEE, pp. 1–6, 2009.

[8] M. Naseriparsa, A.-M. Bidgoli, and T. Varaee, "A Hybrid Feature Selection Method To Improve Performance Of A Group Of Classification Algorithms", *International Journal of Computer Applications*, vol. 69, no. 17, pp. 28-35, 2014.

[9] Y. Danane and T. Parvat, "Intrusion Detection System Using Fuzzy Genetic Algorithm", *In Pervasive Computing (ICPC), International Conference on*, IEEE, pp. 1–5, 2015.

[10] G. Wang, J. Hao, J. Ma, and L. Huang, "A New Approach To Intrusion Detection Using Artificial Neural Networks And Fuzzy Clustering", *Expert systems with applications*, vol. 37, no. 9, pp. 6225–6232, 2010.

[11] A. Feizollah, S. Shamshirband, N. B. Anuar, R. Salleh, and M. L. M. Kiah, "Anomaly Detection Using Cooperative Fuzzy Logic Controller", *In FIRA RoboWorld Congress*, Springer, pp. 220–231, 2013. DOI: 10.1007/978-3-642-40409-2_19.

[12] A. Einipour, "Intelligent Intrusion Detection In Computer Networks Using Fuzzy Systems", *Global Journal of Computer Science and Technology*, vol. 12, no. 11, pp. 19–29, 2012.

[13] J. Amudhavel, V. Brindha, B. Anantharaj, P. Karthikeyan, B. Bhuvaneswari, M. Vasanthi, D. Nivetha, and D. Vinodha, "A Survey On Intrusion Detection System: State Of The Art Review", *Indian Journal of Science and Technology*, vol. 9, no. 11, pp. 1-9, 2016.

[14] S. Axelsson, "Intrusion Detection Systems: A Survey and Taxonomy", *Technical Report, Department of Computer Engineering, Chalmers University*, pp. 15-99, March, 2000.

[15] M. Almseidin, M. Alzubi, S. Kovacs, and M. Alkasassbeh, "Evaluation Of Machine Learning Algorithms For Intrusion Detection System", *In Intelligent Systems and Informatics (SISY), 2017 IEEE 15th International Symposium on*, IEEE, pp. 000277–000282, 2017.11

# ANNEXURE 1

*Table 7: The Obtained Fuzzy Set Parameters Of FRI-IDS Model.*

| Packet Rate | Very Low | Low | Medium | High |
|---|---|---|---|---|
| | [1  1  35.92  91.78] | [166.81  222.66  278.51  334.36] | [475.73  531.58  587.43  643.28] | [950.67  1006.52  1062.37  1118] |
| Byte Rate | Low | Medium | High | |
| | [55  55  83268.03  167136.28] | [461330.38  545198.63  629066.88  712935.13] | [1425835.73  1509703.98  1593572.23  1677420] | |
| Utilization | Low | Medium | High | |
| | [3  3  594.18  11235.33] | [594.18  11235.33  12417.68  23058.83] | [12417.68  23058.83  23650  23650] | |

# ANNEXURE 2

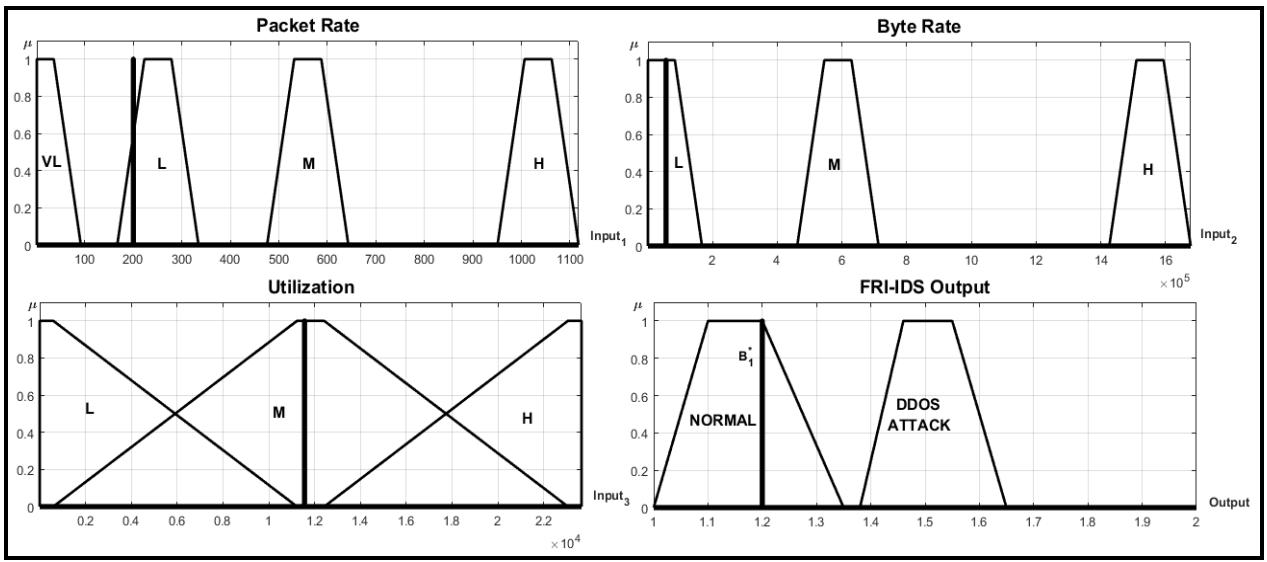

*Figure 5: FRI-IDS Output Response in Case of Normal*

# ANNEXURE 3

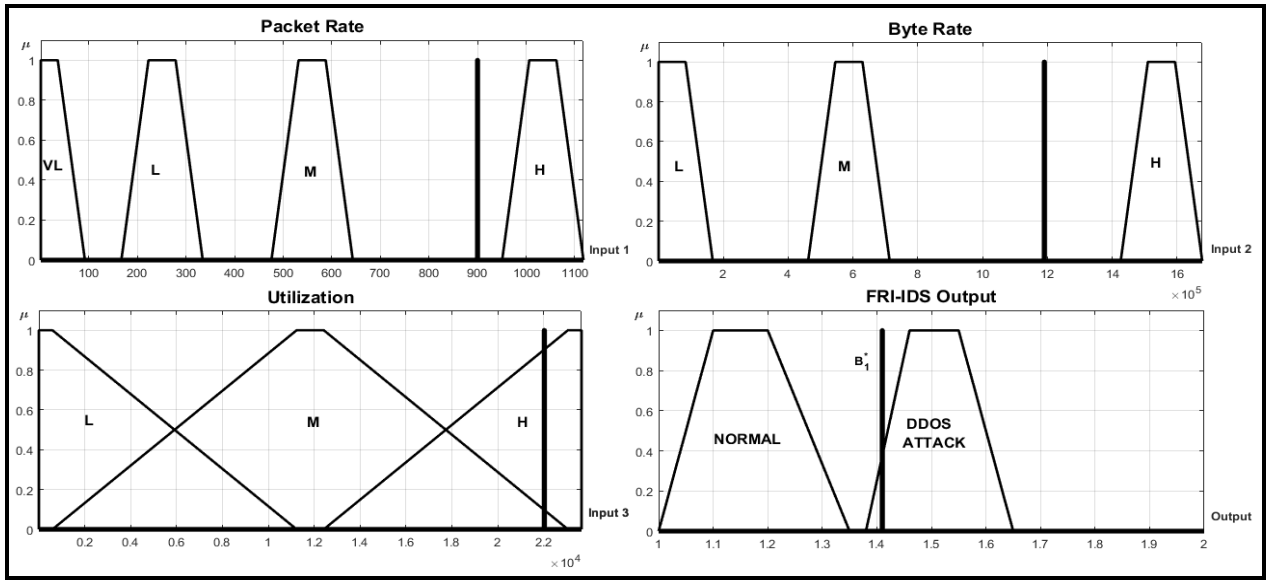

*Figure 6: FRI-IDS Output Response in Case of Attack*